\documentclass[12pt]{article}
\usepackage{csvsimple}
\usepackage{graphicx, setspace, amssymb, subfigure, url, multirow, booktabs, verbatim, bm,threeparttable, amsthm, color}
\usepackage{amsmath} 
\usepackage{indentfirst}
\usepackage{bm}
\usepackage{multirow}
\usepackage[round]{natbib}
\usepackage{appendix}
\usepackage{mathrsfs}
\usepackage{graphicx}
\usepackage{subfigure}
\usepackage[ruled,vlined]{algorithm2e}
\usepackage{caption}
\usepackage{footnote}
\usepackage[figuresright]{rotating}
\usepackage{lscape}
\usepackage{float}
\usepackage{array}
\usepackage{booktabs}
\usepackage{makecell}
\usepackage{enumitem}
\usepackage[colorlinks,linkcolor=blue,anchorcolor=blue,citecolor=blue]{hyperref}
\usepackage{authblk}
\usepackage{CJKutf8}

\theoremstyle{plain}

\newtheorem{theorem}{Theorem}[section]

\newtheorem{condition}{Condition}
\newtheorem{remark}{Remark}
\usepackage[ruled]{algorithm2e}

\def\bse{\begin{eqnarray*}}
\def\ese{\end{eqnarray*}}
\def\be{\begin{eqnarray}}
\def\ee{\end{eqnarray}}
\newcommand{\bit}{\begin{itemize}}
\newcommand{\eit}{\end{itemize}}

\allowdisplaybreaks[3]

\def\boxit#1{\vbox{\hrule\hbox{\vrule\kern6pt  \vbox{\kern6pt#1\kern6pt}\kern6pt\vrule}\hrule}}

\begin{document}
\setlength{\textheight}{650pt}
\setlength{\baselineskip}{20pt}
\title{\bf 
Adaptive Multi-Prior Lasso for High-Dimensional Generalized Linear Models}

\author{
\ Fuzhi Xu\textsuperscript{a},
\ Weijuan Liang\textsuperscript{b},
\ Shuangge Ma\textsuperscript{c}
and 
\ Qingzhao Zhang \textsuperscript{d}\thanks{Correspondence address: qzzhang@xmu.edu.cn}
\\

\textsuperscript{a} School of Management, \\
University of Science and Technology of China, Hefei, China \\
\textsuperscript{b} School of Mathematical Science, Xiamen University, Xiamen, China \\
\textsuperscript{c} Department of Biostatistics, Yale School of Public Health,  New Haven, Connecticut, USA\\
\textsuperscript{d} Department of Statistics and Data Science, School of Economics, and The Wang Yanan Institute for Studies in Economics, Xiamen University, Beijing, China \\
}

\date{}
\maketitle

\begin{abstract}
Incorporation of external information into high-dimensional modeling for gene expression data has been shown, both theoretically and empirically, to substantially enhance performance. Such external information, sometimes referred to as prior information or priors, has become increasingly accessible from multiple sources, yet its reliability may vary considerably. Existing approaches often integrate these priors without sufficiently accounting for their quality, which may result in unsatisfactory or even misleading results.
To effectively and selectively exploit such priors, we propose adaptive Multi-Prior Lasso, a novel regularization approach that simultaneously identifies reliable prior sources and integrates them to improve model performance. For high-dimensional generalized linear models (GLMs), an adaptive data-driven weight is assigned to each prior, so that more reliable sources are emphasized while less credible ones are downweighted. Theoretical guarantees are established, and the proposed method is shown through extensive simulations to improve estimation, prediction, and variable selection. An application to TCGA breast cancer gene expression data further illustrates the practical value of the proposed method, showing that incorporating prior information from PubMed published studies improves model performance.
\end{abstract}


\section{Introduction}
Recent advances have facilitated the collection of large-scale data in a wide range of scientific fields, especially in genetic and genomic studies, where modern platforms routinely generate measurements on tens of thousands to millions of molecular features for each subject. 
For example, in triple-negative breast cancer (TNBC) research, large-scale genomic studies have characterized complex mutational landscapes using hundreds of tumor samples from clinical cohorts, together with transcriptomic measurements on tens of thousands of genes and paired whole-exome sequencing data \cite{yao2025mutational}.
While these studies provide unprecedented opportunities for scientific discovery, they are often characterized by high dimensionality, limited sample sizes, and weak signals, leading to model instability, poor generalizability, and difficulty in identifying truly relevant variables \citep{tseng2015integrating, zhao2015integrative}.
Fortunately, multiple studies have often been conducted on the same scientific problem, providing diverse sources of prior evidence. For instance, several genetic studies on TNBC have identified recurring genetic factors associated with disease progression (e.g., \citealp{alam2022gene}; \citealp{JING2023119542}). Following \citet{yuan2016priorlasso}, we refer to accumulated findings from such studies as ``prior information'' or ``priors'', which represent evidence from previous related studies rather than Bayesian prior distributions. Leveraging such prior information has been shown to substantially improve model performance in high-dimensional settings \citep{wang2023prior, tay2023feature, van2023fast}.

Such prior information is not always reliable. As noted by \citet{yi2022information} and others, retrieved information may not be fully relevant to the current study and may also be incomplete or even misleading. Despite its practical importance, this issue has not been adequately addressed in many existing methods. As illustrated by studies such as \citet{raghu2020pipeline}, incorporating unreliable prior information without proper control can degrade model performance by introducing bias and reducing stability. These considerations underscore the need for methods that can assess the relevance of multiple prior sources and selectively incorporate useful information while limiting the influence of unreliable ones.

\begin{figure}
\centering
	\includegraphics[width=400pt]{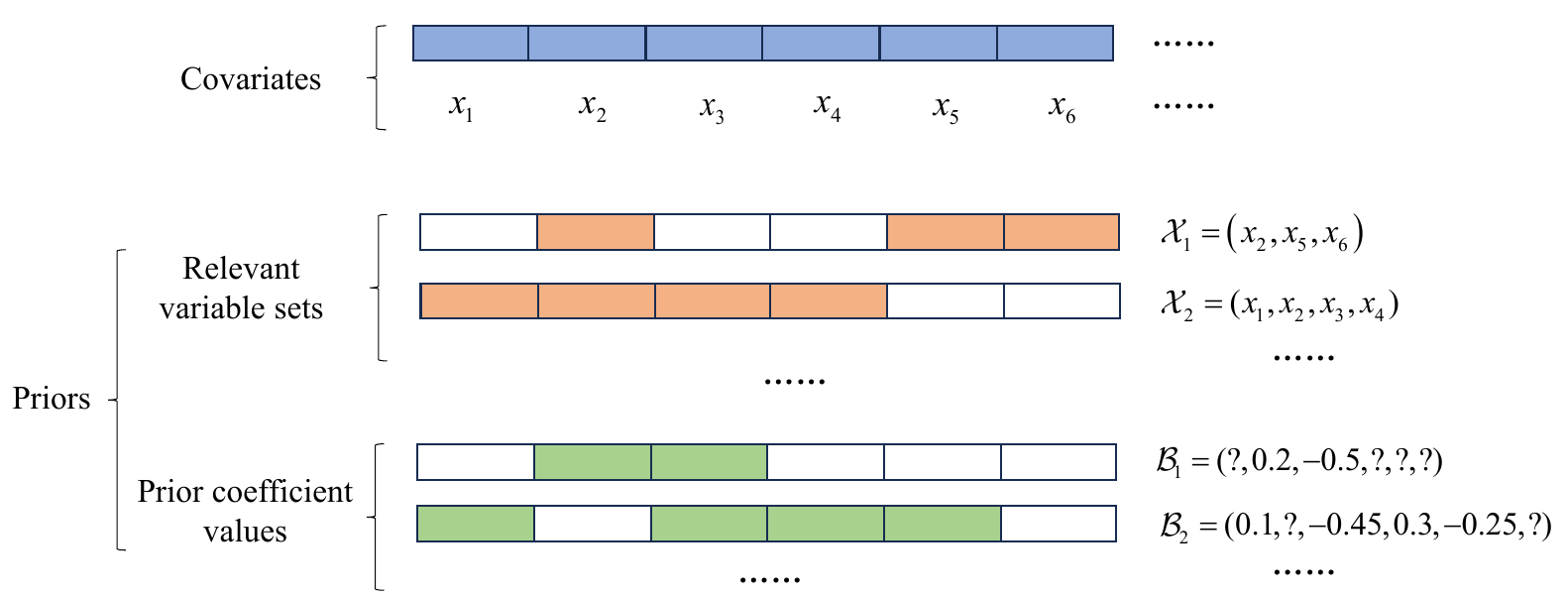}
\caption{Schematic of prior information. $\mathcal{X}$ and $\mathcal{B}$ denote relevant variable sets and prior coefficient values, respectively. Blue = covariates; orange = relevant covariate names; green = coefficient values. Empty (uncolored) cells indicate covariates not marked as relevant by the prior. For example, $\mathcal{X}_1$ flags $x_2$, $x_5$, and $x_6$; $\mathcal{B}_1$ assigns $0.2$ and $-0.5$ to $x_2$ and $x_3$. Question marks in $\mathcal{B}_i$'s indicate missing prior information.}
\label{prior}
\end{figure}

Our goal is to improve high-dimensional modeling for current data by incorporating multiple sources of external information while explicitly accounting for their varying reliability. Rather than treating all priors equally or collapsing them into a single summary, we propose to borrow information adaptively, assigning greater weight to sources that are more compatible with the current data and less weight to those that are less relevant. In particular, we consider two common forms of prior information: (i) \emph{relevant variable sets}, which indicate variables potentially associated with the response, and (ii) \emph{prior coefficient values}, which provide effect-size information from previous studies. These two forms of prior information are illustrated in Figure \ref{prior}.

Penalized estimation methods have demonstrated substantial utilization by incorporating prior information into high-dimensional models. \citet{yuan2016priorlasso} introduced Prior Lasso, which augments the standard Lasso objective with a discrepancy term that encourages agreement between the fitted model and the prior information. This idea has since been extended to various methodological frameworks and applied to diverse studies, such as a quasi-likelihood approach for gene--environment interactions in melanoma and glioblastoma \citep{wang2019identifying}, information-incorporated Gaussian graphical models for lung cancer \citep{yi2022information}, and deep learning-based feature selection in melanoma \citep{wang2023prior, wang2024integrative}. These approaches typically rely on a single prior or treat multiple priors equally, overlooking the issue that not all priors are equally reliable. 

A rigorous framework that can adaptively distinguish and integrate multiple sources of priors is essential for high-dimensional modeling.
Several recent approaches have sought to borrow information from multiple sources, but each entails certain limitations. \citet{tian2023transfer} proposed a transfer learning framework for high-dimensional GLMs that borrows strength from multiple sources, and the limitation is that it requires access to raw data from the source studies. \citet{rauschenberger2023penalized} relaxed this requirement by relying solely on summary-level measures (e.g., signs and importance weights). Both approaches involve a preliminary step to determine which sources are transferable and cannot simultaneously assess and integrate heterogeneous priors within a single model fit.
Some other strategies encode multiple priors through adaptive penalization. For instance, weighted Lasso \citep{bergersen2011weighted} and feature-weighted elastic net \citep{tay2023feature} assigned variable-specific penalty weights based on external prior information. However, these methods cannot directly incorporate prior coefficient values (effect sizes), and their performance may deteriorate when some priors are misleading, since the resulting estimators tend to adapt to irrelevant variables favored by these priors \citep{tay2023feature}. Bayesian approaches have also been proposed to combine priors from multiple studies \citep{zeng2021incorporating, van2023fast}. Yet, they often suffer from high computational cost, complex hyperparameter tuning, and poor scalability in high dimensions \citep{gu2023synthetic}. Overall, there remains a lack of a unified modeling framework that can simultaneously select, weight, and integrate multiple priors of varying reliability in a data-driven manner.

In this paper, we propose an adaptive Multi-Prior Lasso approach to simultaneously and adaptively incorporate multiple sources of prior information into high-dimensional generalized linear models (GLMs). It offers several methodological advantages over existing approaches.
First, by leveraging external information, it alleviates information scarcity typical of high-dimensional, small-sample data, improving modeling performance. 
Second, although conceptually related to Prior Lasso of \citet{yuan2016priorlasso}, it constitutes a non-trivial 
advancement that accommodates multiple priors with varying reliability. It adaptively distinguishes more reliable priors from less reliable ones and offers the potential to identify which sources warrant continued monitoring. In addition, a novel entropy-based penalty is introduced to prevent over-reliance on any single prior source, reducing the risk of bias caused by unreliable priors and fundamentally differentiating our approach from Prior Lasso. Third, unlike integrative analysis that requires pooling multiple datasets or transfer-learning that relies on raw source data, the proposed approach operates solely on summary-level prior information. This feature provides a flexible alternative for incorporating external information from diverse sources without direct access to their raw data. In addition to extensive numerical evaluations, theoretical guarantees are established for the proposed estimator. It is shown that it possesses a weak oracle property and that its data-driven weighting scheme for priors is asymptotically consistent. Furthermore, when truly reliable priors are present, the method automatically assigns them higher weights, effectively prioritizing their information and substantially reducing the risk of being misled by unreliable sources. Finally,
the proposed approach is user-friendly and computationally efficient, offering a practical tool.

The remainder of this paper is organized as follows.
Section \ref{Methodology} introduces the proposed method, establishes its theoretical properties, and presents an efficient computational algorithm. Section \ref{Simulation} reports extensive simulation results.
Section \ref{Real Data Study} applies the method to a TNBC dataset to demonstrate its practical utility. Section \ref{Discussion} concludes with some discussions.
Technical proofs, additional figures, tables, and supplementary numerical results are provided in Supplementary Material.

\section{Methodology}
\label{Methodology}

Assume $n$ i.i.d. samples $\{(\boldsymbol{X}_{i},Y_{i})\}_{i=1}^n$ for the current study, where $\boldsymbol{X}_{i}=(X_{i1},\ldots,X_{ip})^{\top} \in \mathbb{R}^{p}$ denotes the high-dimensional covariate vector (e.g., gene expression measurements) and $Y_i$ denotes the response for the $i$th sample. 
Assume the GLM: 
$$
f(Y_i|\boldsymbol{X}_i) = \exp \{Y_i \theta_i -b(\theta_i) + c(Y_i)\},
$$
where $\theta_i= \boldsymbol{Z}_i^\top \boldsymbol{\beta}$, $\boldsymbol{Z}_i = (1,\boldsymbol{X}_i^\top)^\top$, and $\boldsymbol{\beta}=( \beta_0,\beta_1,...,\beta_p )^{\top}$ denotes a coefficient vector. As commonly assumed, $b(\cdot)$ is twice continuously differentiable with  $b^{\prime\prime}(\cdot) >0$. The loss function, up to an additive function $c(Y_i)$, is given by 
\begin{equation}
\begin{aligned}
\label{glm}
l(\boldsymbol{\beta};\boldsymbol{X},\boldsymbol{Y})
&=-\frac{1}{n} \sum_{i=1}^n[Y_i \theta_i - b(\theta_i)] =-\frac{1}{n} \sum_{i=1}^n [Y_i \boldsymbol{Z}_i^\top \boldsymbol{\beta} - b(\boldsymbol{Z}_i^\top \boldsymbol{\beta})],
\end{aligned}
\end{equation}
where $\boldsymbol{Y}=(Y_1,...,Y_n)^{\top}$ and $\boldsymbol{X}=(\boldsymbol{X}_{1},...,\boldsymbol{X}_{n})^{\top}$.
\cite{yuan2016priorlasso} proposed summarizing prior information into the predicted responses denoted by $\widehat{\boldsymbol{Y}}^p =(\widehat{Y}_1^p, \cdots, \widehat{Y}_n^p)^\top$, where $\widehat{Y}_i^p = b^{\prime}(\boldsymbol{Z}_i^\top \widehat{\boldsymbol{\beta}}^p)$ is the expectation of $Y_i$ given $\boldsymbol{Z}_i$ and a prior estimator $\widehat{\boldsymbol{\beta}}^p$ of $\boldsymbol{\beta}$. 
Depending on the nature of priors, the form of $\widehat{\boldsymbol{\beta}}^p$ may vary. Two common types of prior information are prior coefficient values $\mathcal{B}$ and relevant variable sets $\mathcal{X}$, both retrieved from previous studies. For the former, $\widehat{\boldsymbol{\beta}}^p$ is directly plugged into the calculation of $\widehat{\boldsymbol{Y}}^p$; for the latter, it is defined as:
\begin{equation}\label{lassoest}
    \widehat{\boldsymbol{\beta}}^p =   \mathop{\arg\min}\limits_{\boldsymbol{\beta}\in \mathbb{R}^{p+1} } \bigg\{  -\frac{1}{n} \sum_{i=1}^n [Y_i \boldsymbol{Z}_i^\top \boldsymbol{\beta} - b(\boldsymbol{Z}_i^\top \boldsymbol{\beta})] + \nu \sum_{j:x_j\notin \mathcal{X}} |\beta_j| \bigg\},
\end{equation}
where 
$x_j$ represents the $j$-th covariate,
$\nu$ is a tuning parameter, and this penalized estimation forces covariates in $\mathcal{X}$ to be retained in the model. In this step, the prior information is ``fully trusted'' when obtaining $\widehat{\boldsymbol{\beta}}^p$. To balance the trade-off between the current data and the prior information summarized in $\widehat{\boldsymbol{Y}}^p$, the Prior Lasso estimator is defined as the minimizer of
\begin{equation}\label{priorlasso}
    Q_{\lambda,\eta}(\boldsymbol{\beta}; \boldsymbol{X}, \boldsymbol{Y}, \widehat{\boldsymbol{Y}}^p) = l(\boldsymbol{\beta}; \boldsymbol{X}, \boldsymbol{Y}) + \eta l(\boldsymbol{\beta}; \boldsymbol{X}, \widehat{\boldsymbol{Y}}^p)+\lambda \sum_{j=1}^p |\beta_j|,
\end{equation}
where $\eta$ and $\lambda$ are tuning parameters.

\subsection{Adaptive Multi-Prior Lasso} \label{Multi-Prior Lasso}

Suppose that $M$ priors are available, each consisting of a relevant variable set, coefficient values, or both.
To simultaneously and adaptively identify reliable priors and integrate them into the current model, we propose the adaptive Multi-Prior Lasso objective function:
\begin{align} \label{obj}
Q_{\lambda,\eta,\tau}\big(\boldsymbol{\beta},\boldsymbol{w};\boldsymbol{X},\boldsymbol{Y},\widehat{\boldsymbol{Y}}^{p}\big)
&=l(\boldsymbol{\beta};\boldsymbol{X},\boldsymbol{Y})+\eta\sum_{m=1}^M  w_mD\big(\boldsymbol{\beta};\boldsymbol{X},\widehat{\boldsymbol{Y}}^{p}_m\big) \nonumber\\
&\quad +\tau\sum_{m=1}^Mw_m \log w_m +\lambda\sum_{j=1}^p\vert\beta_j\vert,\nonumber\\
& ~ \text{s.t. } \sum_{m=1}^Mw_m=1, w_m\ge 0,
\end{align}
where $D(\boldsymbol{\beta};\boldsymbol{X},\widehat{\boldsymbol{Y}}^{p}_m)=l(\boldsymbol{\beta};\boldsymbol{X},\widehat{\boldsymbol{Y}}^{p}_m)-l(\widehat{\boldsymbol{\beta}}^{p}_m;\boldsymbol{X},\widehat{\boldsymbol{Y}}^{p}_m)$ denotes the deviance associated with the $m$-th prior. 
Here, $\widehat{\boldsymbol{\beta}}_m^p$ and $\widehat{\boldsymbol{Y}}_m^p$ are obtained by incorporating the prior information from either relevant variable set $\mathcal{X}_m$ or coefficient values $\mathcal{B}_m$, following the Prior Lasso procedure. We adopt the convention $w_m \log w_m = 0$ when $w_m = 0$ to ensure continuity.

The objective function (\ref{obj}) consists of four components.
The first term represents the conventional loss function based on the observed data. 
The second term aggregates deviances $D(\boldsymbol{\beta}; \boldsymbol{X}, \widehat{\boldsymbol{Y}}^p_m)$ across all priors, weighted by adaptively estimated weights $w_m$'s. This term encourages the integration of priors that are more compatible with the current data. 
The third term introduces an entropy penalty on weights $w_m$'s, discouraging over-reliance on any single prior and promoting a more balanced use of multiple sources.
The last term is an $\ell_1$-penalty on $\boldsymbol{\beta}$.
There are three tuning parameters: $\eta, \lambda$ and $\tau$.
$\eta \ge 0$ controls the trade-off between the goodness-of-fit based on the current study and the aggregated prior-based deviance.
Intuitively, a larger $\eta$ is preferred when the priors are more reliable; conversely, a smaller $\eta$ shifts the emphasis toward the current data when the priors are less trustworthy.  $\tau > 0$ controls the strength of the entropy penalty and thus the degree of weight concentration.  $\lambda$ governs the strength of the sparsity-inducing $\ell_1$-penalty.

Based on \eqref{obj}, the adaptive Multi-Prior Lasso estimator can be defined as
$(\widehat{\boldsymbol{\beta}},\widehat{\boldsymbol{w}})=\mathop{\arg\min}_{(\boldsymbol{\beta},\boldsymbol{w})} Q_{\lambda,\eta,\tau}(\boldsymbol{\beta},\boldsymbol{w};\boldsymbol{X},\boldsymbol{Y},\widehat{\boldsymbol{Y}}^{p}).$
Each prior is initially fully trusted in constructing $\widehat{\boldsymbol{\beta}}_m^p$, and its influence is then adaptively adjusted based on its compatibility with the current data.
Notably, in contrast to the use of post-fit negative log-likelihood $l(\boldsymbol{\beta}; \boldsymbol{X}, \widehat{\boldsymbol{Y}}_m^p)$ in \citet{yuan2016priorlasso}, we adopt deviance $D(\boldsymbol{\beta};\boldsymbol{X},\widehat{\boldsymbol{Y}}^{p}_m)$ to better capture the convergence behavior of $\widehat{\boldsymbol{\beta}}^p$ under varying prior sources.

\begin{remark}
To motivate the use of the deviance over the negative log-likelihood in (\ref{obj}), we consider the Gaussian linear model. In this case,
$
l(\widehat{\boldsymbol{\beta}}_m^p;\boldsymbol{X},\widehat{\boldsymbol{Y}}^{p}_m)
=
-\frac{1}{2n}\widehat{\boldsymbol{Y}}_m^{p\top}\widehat{\boldsymbol{Y}}_m^p,
$ and
$
l(\boldsymbol{\beta};\boldsymbol{X},\widehat{\boldsymbol{Y}}^{p}_m)
=
\frac{1}{2n}\|\boldsymbol{Z}(\widehat{\boldsymbol{\beta}}_m^p-\boldsymbol{\beta})\|_2^2
-\frac{1}{2n}\widehat{\boldsymbol{Y}}_m^{p\top}\widehat{\boldsymbol{Y}}_m^p.
$
Consequently,
\[
D(\boldsymbol{\beta};\boldsymbol{X},\widehat{\boldsymbol{Y}}^{p}_m)
=
\frac{1}{2n}\|\boldsymbol{Z}(\widehat{\boldsymbol{\beta}}_m^p-\boldsymbol{\beta})\|_2^2,
\]
which captures the discrepancy between $\widehat{\boldsymbol{\beta}}_m^p$ and $\boldsymbol{\beta}$ and thus reflects how well the prior aligns with the current data.  
Furthermore, for a fixed $\boldsymbol{\beta}$, if the negative log-likelihood is used, the optimal weight admits the closed form
\begin{equation*}
\begin{aligned}
w_m(\boldsymbol{\beta})
&=
\frac{\exp\Big\{-\left(\eta/\tau\right)\, l(\boldsymbol{\beta};\boldsymbol{X},\widehat{\boldsymbol{Y}}^{p}_m)\Big\}}
{\sum_{l=1}^M \exp\Big\{-\left(\eta/\tau\right)\, l(\boldsymbol{\beta};\boldsymbol{X},\widehat{\boldsymbol{Y}}^{p}_l)\Big\}} \\
&=
\frac{\exp\Big\{-\eta/(2n\tau)\|\boldsymbol{Z}(\widehat{\boldsymbol{\beta}}_m^p-\boldsymbol{\beta})\|_2^2\Big\}}
{\sum_{l=1}^M \exp\Big\{-\eta/(2n\tau) \big[
\|\boldsymbol{Z}(\widehat{\boldsymbol{\beta}}_l^p-\boldsymbol{\beta})\|_2^2
-(\widehat{\boldsymbol{Y}}_l^{p\top}\widehat{\boldsymbol{Y}}_l^{p}
-\widehat{\boldsymbol{Y}}_m^{p\top}\widehat{\boldsymbol{Y}}_m^{p})
\big]\Big\}}.
\end{aligned}
\end{equation*}
Hence, $w_m(\boldsymbol{\beta})$ depends not only on the discrepancy term 
$\|\boldsymbol{Z}(\widehat{\boldsymbol{\beta}}_m^p-\boldsymbol{\beta})\|_2^2$, 
but also on an additional offset 
$\widehat{\boldsymbol{Y}}_l^{p\top}\widehat{\boldsymbol{Y}}_l^p-\widehat{\boldsymbol{Y}}_m^{p\top}\widehat{\boldsymbol{Y}}_m^p$, 
which is independent of $\boldsymbol{\beta}$. Such an offset may bias the estimation of $\boldsymbol{w}$ and distort comparisons across priors, especially when the magnitudes of $\widehat{\boldsymbol{Y}}_m^p$ vary substantially.  
In contrast, replacing $l(\boldsymbol{\beta};\boldsymbol{X},\widehat{\boldsymbol{Y}}^{p}_m)$ with the deviance $D(\boldsymbol{\beta};\boldsymbol{X},\widehat{\boldsymbol{Y}}^{p}_m)$ removes this offset and ensures that the weights depend solely on the discrepancy of interest.

This rationale extends beyond the Gaussian linear model setting. In GLMs, the deviance retains its interpretation as a measure of model fit relative to a saturated model, and thus provides a scale-free, model-specific discrepancy between the fitted model and prior-based predictions. It is therefore particularly suitable for evaluating and adaptively weighting multiple priors in high-dimensional settings.
\end{remark}

\textbf{A Toy Example.} 
We consider a toy example under a Gaussian linear model with $n=400$ and $\boldsymbol{\beta} \in \mathbb{R}^{1000}$, where only the first 20 entries are nonzero. Two priors with complementary signal components and equal numbers of relevant and irrelevant variables are constructed: $\mathcal{X}_1=\{x_{1},...,x_{10}\}  \cup \{x_{21},...,x_{30}\}$ and  
$\mathcal{X}_2=\{x_{11},...,x_{30}\}$. Details of the data-generating process are provided in the simulation section.
Single-Prior Lasso, defined in (\ref{single}) and selecting the better-performing prior in a data-driven manner, yields Absolute Model Error (AME) $= 0.919$ and Prediction Mean Squared Error (PMSE) = $1.020$.
The standard Lasso, which does not use prior information, results in AME $ = 1.695$ and PMSE $= 1.079$. In contrast, the Multi-Prior Lasso method assigns weights $0.531$ and $0.469$ to the two priors, leveraging both nearly equally, and achieves AME $= 0.871$ and PMSE $= 1.002$. All three methods successfully recover the true signals, but incorporating prior information reduces false positives (FP = $7$ vs. $12$).

\subsection{Statistical properties} 

To align with existing results \citep{Fan&Lv2011,yuan2016priorlasso}, we slightly abuse notations by omitting intercept $\beta_0$, and define $\boldsymbol{\beta} = (\beta_1, \cdots, \beta_p)^\top \in \mathbb{R}^p$. 
Without loss of generality, let $\boldsymbol{\beta}^{*}=(\boldsymbol{\beta}_S^{*\top},\boldsymbol{\beta}_{S^c}^{*\top})^{\top}$ be the true coefficient vector, where $\boldsymbol{\beta}_S^{*}= (\beta_1^*, \cdots, \beta_s^*)^\top$ are the nonzero entries and $\boldsymbol{\beta}_{S^c}^{*}= \boldsymbol{0}$. Here,  $S \subset \{1,\cdots,p\}$ denotes the index set of relevant variables, $S^c$ denotes its complement, and $s = \vert S \vert$ denotes its cardinality. Let $\boldsymbol{X}_S$ and $\boldsymbol{X}_{S^c}$ be the column subsets of $\boldsymbol{X}$ indexed by $S$ and $S^c$, respectively.
For any vector $\boldsymbol{v}\in \mathbb{R}^d$ and matrix $\boldsymbol{A}\in\mathbb{R}^{l\times m}$, define their infinity-norm as $\| \boldsymbol{v}\|_{\infty} = \max_{1\leq j \leq d}{|v_j|}$ and $\|\boldsymbol{A}\|_\infty = \max_{1\leq i \leq l} \sum_{j=1}^m |a_{ij}|$, respectively.
Denote $d_n=\frac{1}{2}\min_j\{\vert \beta_{j}^{*} \vert :\beta_{j}^{*}\neq0 \}$.
For any $\boldsymbol{\theta}=(\theta_1,\cdots,\theta_n)^{\top}\in \mathbb{R}^n$, define 
$\boldsymbol{\mu}(\boldsymbol{\theta})=[b'(\theta_1),\cdots,b'(\theta_n)]^{\top}$
and $
\boldsymbol{\Sigma}(\boldsymbol{\theta})=\text{diag}[b''(\theta_1),\cdots,b''(\theta_n)]^{\top}
$. 
We use standard asymptotic notations $o(\cdot), O(\cdot), o_p(\cdot)$, and $O_p(\cdot)$ with their usual meanings, and use $a \asymp b$ to denote that two sequences are of the same order.
The following conditions are assumed.

\begin{condition}\label{con1}
	The design matrix $\boldsymbol{X}$ satisfies that 
    \begin{enumerate}[itemindent=1.2em]
        \item[(i)] $\big\|[\boldsymbol{X}_S^{\top}\boldsymbol{\Sigma}(\boldsymbol{X}\boldsymbol{\beta}^{*})\boldsymbol{X}_S ]^{-1} \big\|_{\infty}=O(b_sn^{-1})$, where
        $b_s$ denotes a diverging sequence of positive numbers depending on $s$;
        \item[(ii)] $\big\|\boldsymbol{X}_{S^c}^{\top}\boldsymbol{\Sigma}(\boldsymbol{X}\boldsymbol{\beta}^{*})\boldsymbol{X}_S[\boldsymbol{X}_S^{\top}\boldsymbol{\Sigma}(\boldsymbol{X}\boldsymbol{\beta}^{*})\boldsymbol{X}_S ]^{-1} \big\|_{\infty} \le 1-c_0$, with a constant $0 < c_0 < 1$;
         \item[(iii)] $\mathop{\max}\limits_{\boldsymbol{\beta}_S \in \mathcal{N}_0} \mathop{\max}\limits_{1\le j \le p}\lambda_{\max}[\boldsymbol{X}_S^{\top} \text{diag} \{\vert \boldsymbol{X}_j\vert \circ \vert \boldsymbol{\mu}''(\boldsymbol{X}_S\boldsymbol{\beta}_S) \vert \} \boldsymbol{X}_S]=O(n)$ and \\$\mathop{\max}\limits_{\boldsymbol{\beta}_S \in \mathcal{N}_0} \lambda_{\max}[\boldsymbol{X}_S^{\top} \boldsymbol{\Sigma}(\boldsymbol{X}_S\boldsymbol{\beta}_S) \boldsymbol{X}_S] = O(n)$, where $\mathcal{N}_0=\{\boldsymbol{\beta}_S\in\mathbb{R}^s: \| \boldsymbol{\beta}_S-\boldsymbol{\beta}_S^{*} \|_{\infty} \le d_n\}$, $\boldsymbol{X}_{j}$ is the $j$-th column of ${\boldsymbol{X}}$, and $\circ$ denotes the Hadamard product.
    \end{enumerate}
\end{condition} 

\begin{condition}\label{con2}
	$b_s=o\{\min(n^{1/2-\gamma}\sqrt{\log n}, s^{-1}n^{\gamma}/\log n)\}$  and $d_n \ge n^{-\gamma}\log n$, with some $\gamma \in (0,\frac{1}{2}]$. Additionally, $\lambda=o(b_s^{-1}n^{-\gamma}\log n)$ and $\lambda \gg n^{-\alpha}(\log n)^2$ with $\alpha=\min({\frac{1}{2},2\gamma-\alpha_0})$, $s=O(n^{\alpha_0})$, and $\alpha_0 < 1$, and that $\max_{1\le j\le p}\|\boldsymbol{X}_j \|_{\infty}=o(n^{\alpha}/\log n)$ if the responses are unbounded.
\end{condition} 

\begin{condition}\label{con3}
   For any $\boldsymbol{a} \in \mathbb{R}^n$, there exists a constant $\kappa>0$ such that
	\begin{equation}
	\begin{aligned}\nonumber
		P\left(\vert\boldsymbol{a}^{\top}\boldsymbol{Y} - \boldsymbol{a}^{\top}\boldsymbol{\mu}(\boldsymbol{X}\boldsymbol{\beta}^{*}) \vert > \left\|\boldsymbol{a}  \right\|_2\varepsilon\right) \le 2e^{-\kappa\varepsilon^2},
	\end{aligned}
	\end{equation}	
where $\varepsilon \in (0,\infty)$ for bounded responses and $\varepsilon \in (0,\|\boldsymbol{a}  \|_2/\|\boldsymbol{a} \|_{\infty})$ for unbounded responses.
\end{condition} 

For $m=1,\cdots,M$, define $\delta_m^p=\| \widehat{\boldsymbol{\beta}}^p_m-\boldsymbol{\beta}^{*}\|_2^2$,
which measures the discrepancy between prior-based estimator $\widehat{\boldsymbol{\beta}}_m^p$ and true coefficient $\boldsymbol{\beta}^*$. Accordingly, we partition the priors into:
$$
\mathcal{A} = \big\{m: \delta_m^p = O_p(sn^{-2\gamma}(\log n)^2/(1+\eta)^2) \big\},\quad  \mathcal{A}^c = \big\{1,\cdots, M \big\} \backslash \mathcal{A}.
$$ 
Priors in $\mathcal{A}$ are regarded as more reliable than those in $\mathcal{A}^c$. Denote the cardinality of $\mathcal{A}$ and $\mathcal{A}^c$ as $M_{\mathcal{A}}$ and $M_{\mathcal{A}^c}$, respectively. 

\begin{condition}\label{con4}
For any $m =1,..., M$, define the line segment between $\boldsymbol{\beta}^{*}$ and $ \widehat{\boldsymbol{\beta}}^p_{m}$ by $\mathcal{N}_m = \big\{ \widetilde{\boldsymbol{\beta}}_m^p \in \mathbb{R}^p: \widetilde{\boldsymbol{\beta}}_m^p=\xi\boldsymbol{\beta}^* + (1-\xi)\widehat{\boldsymbol{\beta}}^p_{m},~ \xi \in [0,1]  \big\}$.
Then,
\begin{enumerate}[itemindent=1.2em]
\item[(i)] $\max_{\widetilde{\boldsymbol{\beta}}_m^p \in \mathcal{N}_m} \max_{1\le j \le p}\lambda_{\max}\big[\boldsymbol{X}^{\top} \text{diag} \{\vert \boldsymbol{X}_j\vert \circ \vert \boldsymbol{\mu}''(\boldsymbol{X}\widetilde{\boldsymbol{\beta}}_m^p) \vert \} \boldsymbol{X}\big]=O_p(n);$
\item[(ii)] $\mathop{\max}\limits_{\widetilde{\boldsymbol{\beta}}_m^p \in \mathcal{N}_m}\lambda_{\max}[\boldsymbol{X}^{\top}\boldsymbol{\Sigma}(\boldsymbol{X}\widetilde{\boldsymbol{\beta}}^{p}_m)		\boldsymbol{X}]=O_p(n);$
\item[(iii)]  $ \text{~for any~}  \widetilde{\boldsymbol{\beta}}_m^p\in \mathcal{N}_m, m=1,\cdots,M,(\widehat{\boldsymbol{\beta}}^p_m-\boldsymbol{\beta}^{*} )^{\top}\boldsymbol{X}^{\top}\boldsymbol{\Sigma}(\boldsymbol{X}\widetilde{\boldsymbol{\beta}}_m^p)\boldsymbol{X}(\widehat{\boldsymbol{\beta}}^p_m-\boldsymbol{\beta}^{*}) \asymp n\delta_m^p,$ with probability tending to one.
\end{enumerate}

\end{condition}

\begin{condition}\label{con5}
For the priors in $\mathcal{A}$, 
\begin{enumerate}[itemindent=1.2em]
\item[(i)] 
$ \mathop{\max}\limits_{m \in \mathcal{A}} \delta_m^p = o_p(\lambda\eta^{-1});$
\item[(ii)] $\mathop{\max} \limits_{m \in \mathcal{A}} \big\|\boldsymbol{X}^{\top}\boldsymbol{\Sigma}(\boldsymbol{X}\boldsymbol{\beta}^{*})\boldsymbol{X} (\widehat{\boldsymbol{\beta}}^p_m-\boldsymbol{\beta}^{*}) \big\|_{\infty} \le C_1n\lambda\eta^{-1},$ with probability tending to one, for some constant $C_1>0$.
\end{enumerate}
\end{condition}

\begin{condition}\label{con6}
For the priors in $\mathcal{A}^c$, 
\begin{enumerate} [itemindent=1.2em]
\item[(i)]  
$M_{\mathcal{A}}^{-1}M_{\mathcal{A}^c}\exp \Big\{-({\eta}/{\tau})\min\limits_{m\in \mathcal{A}^c}\delta_m^p   \Big\} =o_p(1);$
\item[(ii)]
there exists a constant $C_2>0$ such that, with probability tending to one,
\begin{equation*}
\begin{aligned}
& M_{\mathcal{A}}^{-1}M_{\mathcal{A}^c}\exp \Big\{-({\eta}/{\tau})\min_{m\in \mathcal{A}^c} \delta_m^p  \Big\} 
\Big\| \boldsymbol{X}^{\top}\boldsymbol{\Sigma}(\boldsymbol{X}\boldsymbol{\beta}^{*})\boldsymbol{X}(\widehat{\boldsymbol{\beta}}^p_m-\boldsymbol{\beta}^{*}) \Big\|_{\infty} \\
& ~ \le   C_2 n \lambda \eta^{-1},~~ \text{for~} m \in \mathcal{A}^c;    
\end{aligned}    
\end{equation*}
\item[(iii)] $\widehat{\boldsymbol{\beta}}_{m}^p$ satisfies $$M_{\mathcal{A}}^{-1}M_{\mathcal{A}^c}\exp \Big\{-({\eta}/{\tau})\min_{m\in \mathcal{A}^c} \delta_m^p  \Big\} \delta_m^p
=o_p(\lambda\eta^{-1}),~~ \text{for~} m \in \mathcal{A}^c.$$
\end{enumerate}
\end{condition}

Conditions \ref{con1}-\ref{con3} reformulate the covariate-design and tail-bound requirements of existing studies \citep{Fan&Lv2011,yuan2016priorlasso}. Condition \ref{con4} adapts Condition D of \citet{yuan2016priorlasso}. 
Conditions \ref{con5}-\ref{con6} specify the technical requirements on multiple priors that are needed to establish the weak-oracle property and consistency of the data-driven weights. In particular, Condition \ref{con5} requires that reliable priors in $\mathcal{A}$ yield prior estimators sufficiently close to the true coefficient vector, whereas Condition \ref{con6} ensures that deviances for priors in $\mathcal{A}^c$ are sufficiently larger than that for priors in $\mathcal{A}$, making reliable priors distinguishable from unreliable ones. We note that either $\mathcal{A}$ or $\mathcal{A}^c$ may be empty. If $\mathcal{A} = \emptyset$, $\widehat{\boldsymbol{\beta}}$ reduces to the standard Lasso-penalized estimator, and we address this case in Appendix A  in Supplementary Material. Conversely, if $\mathcal{A}^c=\emptyset$, Theorem \ref{thm1} holds without invoking Condition \ref{con6}.




\begin{theorem}\label{thm1}
  Assume Conditions \ref{con1}-\ref{con6} hold. Suppose ${\eta}/{\tau}sn^{-2\gamma}(\log n)^2/(1+\eta)^2\leq C$ and $\log p=O(n^{1-2\alpha})$, where $C>0$ is a constant. Then, there exists a minimizer $\widehat{\boldsymbol{\beta}} = (\widehat{\boldsymbol{\beta}}_S^\top, \widehat{\boldsymbol{\beta}}_{S^c}^\top)^\top$ of $Q_{\lambda, \eta, \tau} (\boldsymbol{\beta}, \boldsymbol{w}; \boldsymbol{X}, \boldsymbol{Y}, \widehat{\boldsymbol{Y}}^p)$ such that
    \begin{equation}
	\begin{aligned}\label{thm-1}
		\widehat{\boldsymbol{\beta}}_{S^c}=\boldsymbol{0},
	\end{aligned}
\end{equation}
  with probability tending to one, and 
\begin{equation}
	\begin{aligned}\label{thm-2}
		\big\|\boldsymbol{\widehat{\beta}}_S-\boldsymbol{\beta}_S^{*} \big\|_{\infty}= O_p(n^{-\gamma}\log n/(1+\eta)).
	\end{aligned}
\end{equation}
	 In addition, the estimated weights satisfy
\begin{equation}
	\begin{aligned}\label{thm-3}
		\sum_{m \in \mathcal{A}}\widehat{w}_m {\rightarrow} 1, ~~\text{and}~~\widehat{w}_m {\asymp}  \frac{1}{M_{\mathcal{A}}} ~ \text{for all}~ m\in \mathcal{A},
	\end{aligned}
\end{equation} 
with probability tending to one.
\end{theorem}

Theorem \ref{thm1} (\ref{thm-1}) shows that adaptive Multi-Prior Lasso achieves variable-selection consistency comparable to that of both Lasso and  Prior Lasso. When at least one prior is reliable (i.e., $\mathcal{A}\neq\emptyset$), the $\ell_\infty$‐norm bound for $\widehat{\boldsymbol{\beta}}_S$ established in (\ref{thm-2}) attains order $O(n^{-\gamma}\log n/(1+\eta))$, which improves upon the Lasso rate $O(n^{-\gamma}\log n)$ and matches that of Prior Lasso up to factor $(1+\eta)^{-1}$.
Tuning parameter $\eta$ thus controls the continuum between these extremes: $\eta = 0$  recovers the Lasso estimator, whereas $\eta\rightarrow \infty$  yields strictly faster convergence than Lasso. 
Moreover, (\ref{thm-3}) indicates that, with high probability, the proposed method concentrates its weights on all reliable priors, thereby improving estimation accuracy. Furthermore, when multiple reliable priors are present, their estimated weights have the same order, ensuring that all such priors contribute comparably to $\widehat{\boldsymbol{\beta}}$. This property is guaranteed by the entropy penalty, which prevents over-concentration on any single prior source.

\subsection{Computation} \label{Computation}

We estimate $\boldsymbol{\beta}$ and $\boldsymbol{w}$ using an iterative scheme summarized in Algorithm \ref{algorithm1}. $Q_{\lambda,\eta,\tau}(\boldsymbol{\beta},\boldsymbol{w})$ is convex in 
$\boldsymbol{\beta}$ when $\boldsymbol{w}$ is fixed and strictly convex in $\boldsymbol{w}$ when $\boldsymbol{\beta}$ is fixed. Moreover, the weight-update subproblem admits a unique closed-form solution, while the $\boldsymbol{\beta}$-update reduces to a Prior Lasso problem. These properties allow the joint minimization to be efficiently carried out through alternating updates.

\begin{algorithm}[h!] 
\SetAlgoLined
\caption{Iterative Estimation Procedure}
\label{algorithm1}
\KwIn{Data $\{(\boldsymbol{X}_{i},Y_{i})\}_{i=1}^n$; tuning parameters $\eta$, $\tau$, and $\lambda$; initial value $\boldsymbol{\beta}^{(0)}$.}
\KwOut{Estimated coefficient vector $\widehat{\boldsymbol{\beta}}$ and weight vector $\widehat{\boldsymbol{w}}$.}
\For{$t = 1, 2, \cdots$ until convergence}{
    \textbf{Step 1:} Update $\boldsymbol{w}^{(t)}$ based on current $\boldsymbol{\beta}^{(t-1)}$ as:
    \begin{equation}
    \begin{aligned}\label{w}
     {\boldsymbol{w}}^{(t)}  = \mathop{\arg\min}\limits_{\boldsymbol{w}} Q_{\lambda,\eta,\tau}\big({\boldsymbol{\beta}}^{(t-1)},\boldsymbol{w};\boldsymbol{X},\boldsymbol{Y},\widehat{\boldsymbol{Y}}^{p}\big).
    \end{aligned}
    \end{equation}
    \\
    \textbf{Step 2:} Update $\boldsymbol{\beta}^{(t)}$ by minimizing objective function in (\ref{obj}) with fixed $\boldsymbol{w}^{(t)}$ as: 
   \begin{equation}
    \begin{aligned}\label{beta}
     {\boldsymbol{\beta}}^{(t)} = \mathop{\arg\min}\limits_{\boldsymbol{\beta}} Q_{\lambda,\eta,\tau}\big(\boldsymbol{\beta},{\boldsymbol{w}}^{(t)};\boldsymbol{X},\boldsymbol{Y},\widehat{\boldsymbol{Y}}^{p}\big).
    \end{aligned}
   \end{equation}
}
\Return Final estimators $\widehat{\boldsymbol{\beta}}$, $\widehat{\boldsymbol{w}}$.
\end{algorithm}

With $\boldsymbol{\beta}={\boldsymbol{\beta}}^{(t-1)}$ fixed, the minimization problem in (\ref{w}) is equivalent to minimizing $f_{\eta,\tau}(\boldsymbol{w};\boldsymbol{X},\widehat{\boldsymbol{Y}}^p)$ with respect to $\boldsymbol{w}$, where
\begin{equation}
\begin{aligned}\nonumber
f_{\eta,\tau}\big(\boldsymbol{w};\boldsymbol{X},\widehat{\boldsymbol{Y}}^p\big)&=\eta\sum_{m=1}^Mw_mD\big(\boldsymbol{\beta}^{(t-1)};\boldsymbol{X},\widehat{\boldsymbol{Y}}^{p}_m\big)+\tau\sum_{m=1}^Mw_m \log w_m,  \\
& \text{s.t.}~\sum_{m=1}^Mw_m =1,~ w_m\ge0.
\end{aligned}
\end{equation}
We solve the constrained problem using the method of Lagrange multipliers, with the Lagrangian function defined as:
\begin{equation}
\begin{aligned}\nonumber
F_{\eta,\tau}\big(\boldsymbol{w};\boldsymbol{X},\widehat{\boldsymbol{Y}}^p\big)= & \eta\sum_{m=1}^Mw_mD\big(\boldsymbol{\beta}^{(t-1)};\boldsymbol{X},\widehat{\boldsymbol{Y}}^{p}_m\big)+\tau\sum_{m=1}^Mw_m\log w_m\\
& -\zeta\Big(\sum_{m=1}^Mw_m-1\Big),
\end{aligned}
\end{equation}
where $\zeta \neq 0$ is the Lagrange multiplier.
Based on the Karush-Kuhn-Tucker (KKT) conditions, we obtain that the $m$-th element of $\boldsymbol{w}^{(t)}$ is
\begin{equation}
\begin{aligned}
\label{weightest}
{w}_m^{(t)} = \frac{\exp\left\{-\left({\eta}/{\tau}\right) D\big(\boldsymbol{\beta}^{(t-1)};\boldsymbol{X},\widehat{\boldsymbol{Y}}^{p}_m\big)\right\}}{\sum_{l=1}^M \exp\left\{-\left({\eta}/{\tau}\right) D\big(\boldsymbol{\beta}^{(t-1)};\boldsymbol{X},\widehat{\boldsymbol{Y}}^{p}_l\big)\right\}}. 
\end{aligned}
\end{equation}
For computational convenience and to reduce tuning burden, we fix $\tau$ via  ${\eta}/{\tau}D_{\min}=C_0$, whenever $\eta \neq 0$ and $D_{\min} \neq 0$, where $D_{\min}=\min_{1\le m \le M}D(\boldsymbol{\beta};\boldsymbol{X}, \widehat{\boldsymbol{Y}}_m^{p})$ and $C_0>0$ is a constant. With this setting, priors whose deviance is of higher order than  $D_{\min}$ receive negligible weights. If $\eta\neq 0$ but $D_{\min} = 0$, we assign equal weights to all zero-deviance priors, setting $w^{(t)}_m = 1/\widetilde{M}$, where $\widetilde{M}$ is the number of priors satisfying $D(\boldsymbol{\beta};\boldsymbol{X}, \widehat{\boldsymbol{Y}}_m^{p}) = 0$.

With $\boldsymbol{w} = \boldsymbol{w}^{(t)}$ fixed, the minimization problem in (\ref{beta}) is equivalent to minimizing $l_{\lambda,\eta}(\boldsymbol{\beta}; \boldsymbol{X}, \boldsymbol{Y}, \widehat{\boldsymbol{Y}}^p)$ with respect to $\boldsymbol{\beta}$, where
\begin{equation}
\begin{aligned}\nonumber
l_{\lambda,\eta}\big(\boldsymbol{\beta};\boldsymbol{X},\boldsymbol{Y},\widehat{\boldsymbol{Y}}^p\big)
= l\left(\boldsymbol{\beta};\boldsymbol{X},\boldsymbol{Y}\right)+ 
\eta\sum_{m=1}^Mw_m^{(t)}l\big(\boldsymbol{\beta};\boldsymbol{X},\widehat{\boldsymbol{Y}}^{p}_m\big) + \lambda\sum_{j=1}^p\vert\beta_j\vert.  
\end{aligned}
\end{equation}
Note that
\begin{align}\label{lasso}
&l_{\lambda,\eta}\big(\boldsymbol{\beta};\boldsymbol{X},\boldsymbol{Y},\widehat{\boldsymbol{Y}}^p\big) \nonumber \\ 
=&-\frac{1}{n}\sum_{i=1}^n\left[\bigg(Y_i+\eta\sum_{m=1}^Mw_m^{(t)}\widehat{Y}_{m,i}^{p}\bigg)\boldsymbol{Z}_i^{\top}\boldsymbol{\beta}-(1+\eta)b\left(\boldsymbol{Z}_i^{\top}\boldsymbol{\beta}\right)  \right]+\lambda\sum_{j=1}^p\vert\beta_j\vert     \nonumber \\ 
\propto&-\frac{1}{n}\left[\frac{Y_i+\eta\sum_{m=1}^Mw_m^{(t)}\widehat{Y}_{m,i}^{p}}{1+\eta}\boldsymbol{Z}_i^{\top}\boldsymbol{\beta}- b\left(\boldsymbol{Z}_i^{\top}\boldsymbol{\beta}\right)  \right] + \frac{\lambda}{1+\eta}\sum_{j=1}^p \vert\beta_j\vert     \nonumber \\ 
= &l\big(\boldsymbol{\beta};\boldsymbol{X},\widetilde{\boldsymbol{Y}}^{(t)}\big)+ \frac{\lambda}{1+\eta}\sum_{j=1}^p \vert\beta_j\vert,
\end{align}
where $\widetilde{\boldsymbol{Y}}^{(t)}= ({\boldsymbol{Y}+\eta\sum_{m=1}^M w_m^{(t)} \widehat{\boldsymbol{Y}}^p_m})/(1+\eta)$ is the adjusted response values incorporating prior information. This formulation takes the same form as the penalized likelihood in GLMs with an $l_1$-penalty. Therefore, any standard Lasso fitting algorithm can be directly applied.

\section{Simulation} \label{Simulation}

We consider two widely used GLMs: linear and logistic regression. In all settings, $p = 1000$, $\boldsymbol{X}$ has independent rows, each generated from a multivariate normal distribution $N(0,\boldsymbol{\Sigma})$, where $\boldsymbol{\Sigma}=\{\sigma_{ij}\}_{i,j=1}^p$ with $\sigma_{ij}=\rho^{\vert i-j\vert}$ and $\rho=0.5$. 
All reported results are averaged over 100 independent replications. In each replication, we independently generate a training set, a validation set (both with sample sizes $n=200$ or $400$), and a testing set of size $n_0 = 500$.

\noindent\textbf{Linear Regression.} $Y_i$ is independently generated from $Y_i\sim N(\boldsymbol{Z}_i^\top\boldsymbol{\beta},1)$, and the true coefficient vector is $\boldsymbol{\beta}= (0, 2, -1.5, -0.5, -2, 0.5, 2, -1.5, 2, -2, 1, 1.5, -2, 1, 1.5, -0.5, -2, 0.5, $ $2, -1.5, 1, 0, \cdots, 0)^{\top}$, with the first 20 covariates except for the intercept term relevant to the response.

\noindent\textbf{Logistic Regression.} $Y_i$ is independently drawn from
$Y_i\sim \text{Bernoulli} (p(\boldsymbol{Z}_i^\top\boldsymbol{\beta}))$, where $  p(\boldsymbol{Z}_{i}^{\top}\boldsymbol{\beta})={\exp{(\boldsymbol{Z}_{i}^{\top}\boldsymbol{\beta})}}/[{1+\exp{(\boldsymbol{Z}_{i}^{\top}\boldsymbol{\beta})}}]$
with true coefficient vector $\boldsymbol{\beta}= (0, -0.5, -1.25, $ $ 0.5, -0.45, 1, 0.75, -2, 1, 1.5, 0.8, 0, \cdots, 0)^{\top},$ and the first 10 covariates except for the intercept term being relevant.

We consider 16 prior scenarios by crossing: 
(1) two model types (linear: $\boldsymbol{S}_1$-$\boldsymbol{S}_8$; logistic: $\boldsymbol{S}_9$-$\boldsymbol{S}_{16}$);
(2) two types of prior information (relevant variable sets: $\boldsymbol{S}_1$-$\boldsymbol{S}_4$, $\boldsymbol{S}_9$-$\boldsymbol{S}_{12}$; coefficient values: $\boldsymbol{S}_5$-$\boldsymbol{S}_8$, $\boldsymbol{S}_{13}$-$\boldsymbol{S}_{16}$); and
(3) four levels of prior reliability (proportion of truly relevant variables or coefficients):
fully reliable ($\boldsymbol{S}_1$, $\boldsymbol{S}_9$) versus full‐set noisy prior ($\boldsymbol{S}_5$, $\boldsymbol{S}_{13}$),
completely unreliable  ($\boldsymbol{S}_2$, $\boldsymbol{S}_{10}$) versus partial-set noisy prior ($\boldsymbol{S}_6$, $\boldsymbol{S}_{14}$),
mixed-quality (4 sources) ($\boldsymbol{S}_3$, $\boldsymbol{S}_{11}$) versus mixed ``full + partial'' noisy (4 sources) ($\boldsymbol{S}_7$, $\boldsymbol{S}_{15}$),
mixed-quality (8 sources) ($\boldsymbol{S}_4$, $\boldsymbol{S}_{12}$) versus mixed ``full + partial'' noisy (8 sources) ($\boldsymbol{S}_8$, $\boldsymbol{S}_{16}$).
See Appendix B in Supplementary Material for details.

We consider the following alternatives:
(1) Single-Prior Lasso (SPL): excludes the entropy penalty and selects a single best prior by minimizing
\begin{align} \label{single}
Q_{\lambda,\eta}\big(\boldsymbol{\beta},\boldsymbol{w};\boldsymbol{X},\boldsymbol{Y},\widehat{\boldsymbol{Y}}^{p}\big)
&=l(\boldsymbol{\beta};\boldsymbol{X},\boldsymbol{Y})+\eta\sum_{m=1}^M w_m l\big(\boldsymbol{\beta};\boldsymbol{X},\widehat{\boldsymbol{Y}}^{p}_m\big) +\lambda\sum_{j=1}^p\vert\beta_j\vert, \nonumber\\
& \text{s.t.}~\sum_{m=1}^Mw_m=1,~ {w_m \ge 0},
\end{align}
where exactly one $w_m = 1$, indicating the selected prior.
(2) Equal-Weight Prior Lasso (EWPL): assigns equal weights to all priors in  (\ref{obj}), without adaptive weighting.
(3) Best-Prior Lasso (BPL): implements the Prior Lasso estimator in (\ref{priorlasso}) using an oracle-specified best prior.
(4) Worst-Prior Lasso (WPL): same as BPL but uses an oracle-specified worst prior.
(5) Best Prior Only (BP): fully trusts the given best prior. For relevant variable set priors, the estimate is obtained via (\ref{lassoest}); for coefficient value priors, the provided vector is directly applied.
(6) Worst Prior Only (WP): analogous to BP but uses the worst prior.
(7) Lasso: standard Lasso without incorporating prior information.
(8) Fwelnet: the feature-weighted elastic net of \cite{tay2023feature}, which incorporates prior information by embedding prior-derived weights into the elastic-net penalty for each variable.
(9) Oracle: fits the model using the true relevant variable set, without performing variable selection.
Methods (3)–(6) depend on oracle knowledge of prior quality, and method (9) relies on oracle true relevant variable sets information. Although these five oracle methods are not feasible in practice, they serve as useful benchmarks.

\begin{figure}
	\centering 
	\includegraphics[width=410pt]{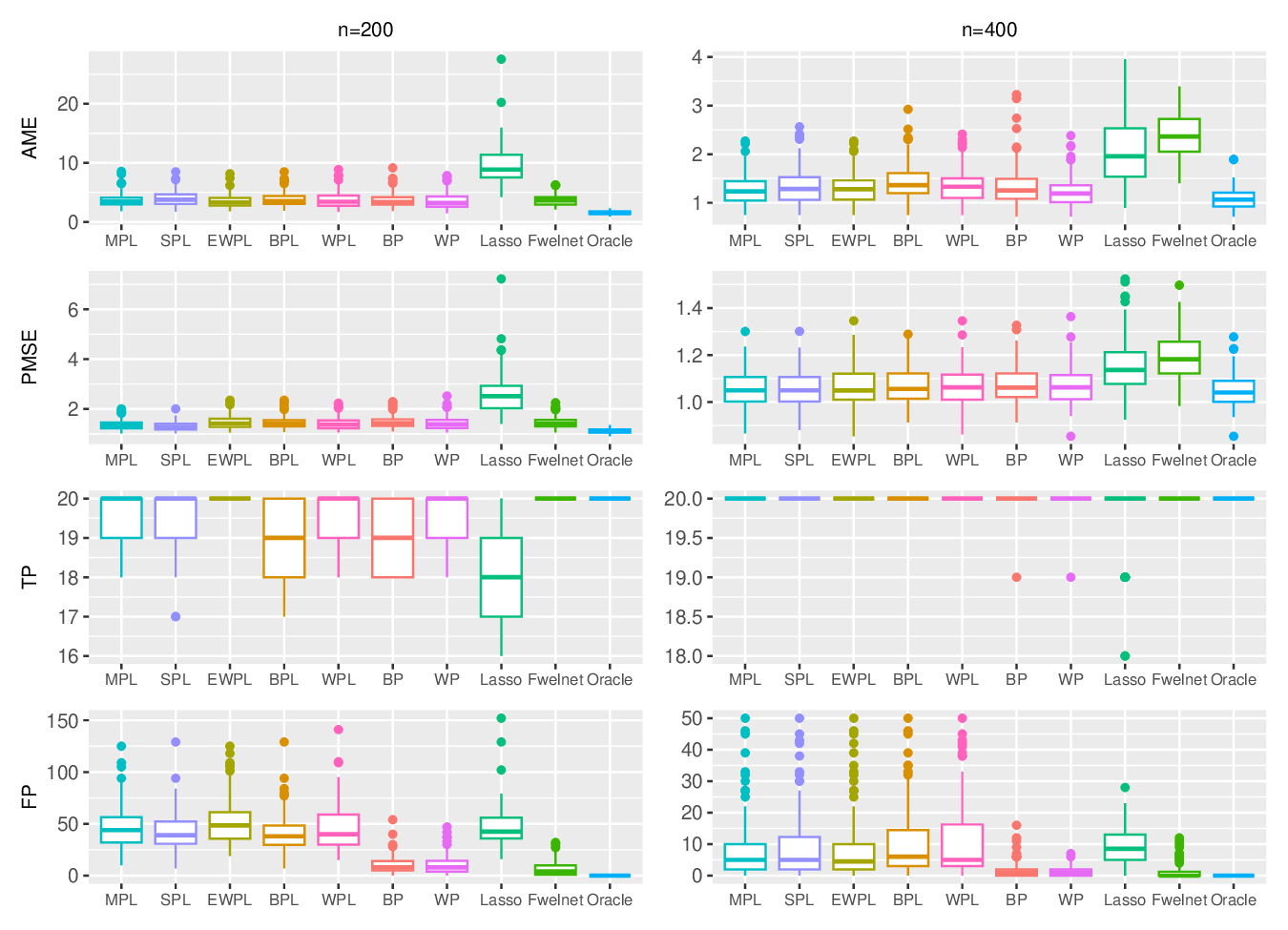}
	\caption{Simulation results for linear regression under scenario $\boldsymbol{S}_1$.} 
        \label{linear_1_5}
\end{figure}
\begin{figure}
	\centering 
	\includegraphics[width=300pt]{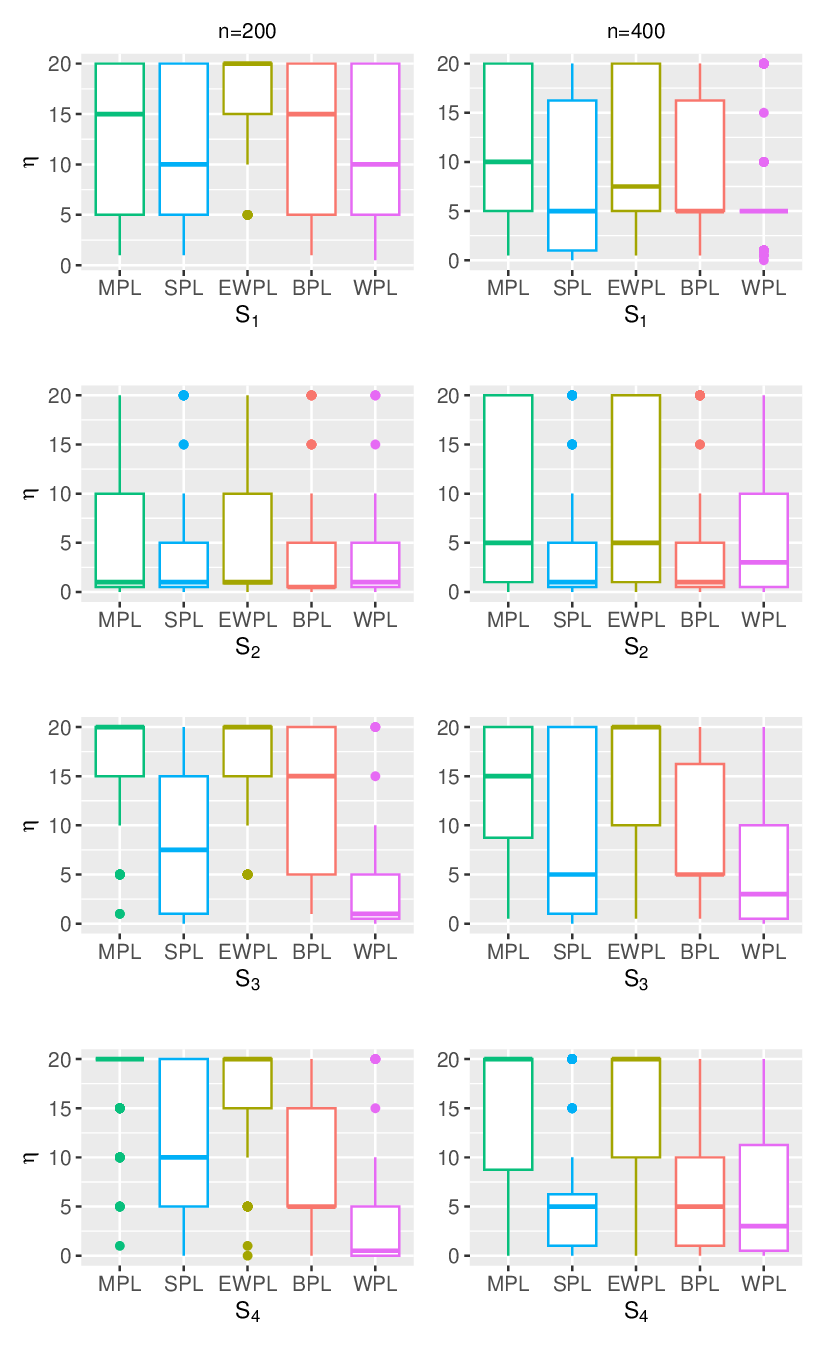}
	\caption{Selection of tuning parameter $\eta$ in linear regression with relevant variable set priors. 
    } 
        \label{linear_1_eta}
\end{figure}

We implement the iterative algorithm described in Section~\ref{Computation}, where the $\ell_1$-penalized GLM step in (\ref{lasso}) is solved using the \texttt{glmnet} package in \textsf{R}. 
For the Fwelnet method, we use the \texttt{fwelnet} R package with its default settings.
Tuning parameters $\lambda$ and $\eta$ are selected by maximizing the average empirical likelihood on a validation set, with $\eta$ chosen from the grid $(0, 0.5, 1, 5, 10, 15, 20)$. When $\eta \neq 0$, we set $C_0 = 1$ (without loss of generality) to determine $\tau$. For linear regression, we assess performance using the following metrics: (1) Absolute Model Error (AME): $\|\widehat{\boldsymbol{\beta}} - \boldsymbol{\beta}\|_1$;
(2) Prediction Mean Squared Error (PMSE): $\|\widehat{\boldsymbol{Y}}^{\text{test}} - \boldsymbol{Y}^{\text{test}}\|_2^2 / n_0$, where $\widehat{Y}_i^{\text{test}} = b'(\boldsymbol{Z}_i^{\text{test}\top} \widehat{\boldsymbol{\beta}})$ denotes the predicted value on the testing set;
(3) True Positives (TP): number of correctly identified nonzero coefficients;
(4) False Positives (FP): number of incorrectly identified nonzero coefficients.
Figure \ref{linear_1_5} provides the boxplots of criteria (1)-(4) for linear regression. Figure \ref{linear_1_eta} reports the optimal selection of $\eta$ for linear regression with relevant variable sets type priors. Additional results are available in Tables B1-B2 and  Figures B1-B18.

Simulation results demonstrate the advantages and robustness of the proposed Multi-Prior Lasso (MPL). When relevant variable sets are fully reliable ($\boldsymbol{S}_1, \boldsymbol{S}_9$) or full‐set noisy coefficients are provided ($\boldsymbol{S}_5, \boldsymbol{S}_{13}$), all prior-guided methods (BP and WP) and prior-incorporated methods (MPL, SPL, EWPL, BPL, and WPL) outperform Lasso, with MPL and EWPL achieving the best overall performance in estimation (lowest AME and ME), prediction (lowest PMSE and PMR), and variable selection (highest TP and lowest FP). The benefits are most evident when relevant variable set priors are used (Figure~\ref{linear_1_5}, Figure~B4). When coefficient value priors are available, the advantage of integrating multiple sources becomes even more pronounced, as Oracle relies only on support information and needs to estimate effect sizes, resulting in slightly inferior performance compared to MPL (Figure~B4, Figure~B13).
For the feature‐weighted elastic net (Fwelnet), performance is comparable to MPL when priors are relevant variable sets but deteriorates under coefficient‐value priors, as Fwelnet can only exploit relevant variable set information. Moreover, Fwelnet shows weaker performance in logistic regression, exhibiting larger AME, ME, and PMR and lower TP than MPL.

When relevant variable sets are completely unreliable ($\boldsymbol{S}_2, \boldsymbol{S}_{10}$) or only partial‐set noisy priors are available ($\boldsymbol{S}_6, \boldsymbol{S}_{14}$), the prior-guided methods (BP and WP), perform the worst by selecting entirely irrelevant variables. MPL, SPL, and EWPL adaptively downweight misleading sources and maintain performance comparable to Lasso, indicating robustness to poor prior quality.
Fwelnet exhibits poorer variable selection and prediction performance in this setting, consistent with the observation that performance degrades markedly when priors are unreliable.



For the scenarios involving mixed quality of relevant variable sets ($\boldsymbol{S}_3$, $\boldsymbol{S}_4$, $\boldsymbol{S}_{11}$, $\boldsymbol{S}_{12}$) or mixed “full + partial” noisy coefficient ($\boldsymbol{S}_7$, $\boldsymbol{S}_8$, $\boldsymbol{S}_{15}$, $\boldsymbol{S}_{16}$), BPL and BP benefit from oracle-specified best priors and perform well by avoiding contamination (Figures B2, B6, B11 and B15). MPL and SPL offer comparable results without oracle knowledge. EWPL, WPL, and WP are more affected by unreliable priors, yet still outperform Lasso. In the presence of coefficient value priors, MPL continues to outperform Oracle by effectively borrowing strength from informative effect sizes. Fwelnet performs comparably to Lasso, as it cannot discount the negative influence of unreliable priors.
We also observe that MPL, SPL, EWPL, and Lasso tend to yield higher FPs than Fwelnet under comparable TPs, which may be attributed to the tendency of Lasso-type methods to over-select variables relative to elastic net. Nevertheless, MPL achieves smaller PMSE and AME overall, indicating improved estimation accuracy when incorporating prior information. This observation is consistent with our theoretical results.


The adaptivity of MPL is further supported by its estimated prior weights, which concentrate on reliable sources when available (Tables B1-B2). Larger values of $\eta$ are selected when the priors contain more relevant variables, allowing the model to place greater emphasis on informative and reliable sources. As the sample size increases, TP increases while FP, AME, ME, and PMSE decrease.

\section{Data Analysis}\label{Real Data Study}
TNBC is a biologically distinct and clinically aggressive subtype of breast cancer, characterized by the negative expression of estrogen receptor (ER), progesterone receptor (PR), and HER-2 amplification on the surface of cancer cells. Compared to other subtypes, TNBC exhibits markedly different risk factors, molecular profiles, and therapeutic responses, including high proliferative activity, immune infiltration, and DNA repair deficiencies. Due to the absence of targeted therapies and limited treatment options, TNBC is often associated with a poor prognosis. 
To investigate genetic factors associated with TNBC, we consider TNBC status as a binary outcome (1 = TNBC, 0 = non-TNBC), with gene expression levels serving as high-dimensional covariates. 
The current dataset is the Breast Invasive Carcinoma cohort (TCGA, Firehose Legacy, \url{https://www.cbioportal.org/study/summary?id=brca\_tcga}), 
comprising 911 samples and 20,210 genes after excluding samples with missing values. Of these, $12.73\%$ are classified as TNBC.

Breast cancer, particularly including TNBC, has been extensively studied, leading to a substantial body of prior knowledge. We collect prior information from six recently published papers \citep{alam2022gene, JING2023119542, huang2021prognostic, chen2022copy, harris2021tumor, huang2021defining} indexed in PubMed, see Table C1 for details. 
{These studies can be regarded as relatively ``good'' priors as they were all published in reputable journals. In terms of scientific relevance, some are closely aligned with the current study, whereas others address related but more distant topics, leading to heterogeneous reliability across priors. Moreover, all six papers were published in recent years, reflecting up-to-date experimental platforms and analytical techniques and thereby reducing discrepancies due to outdated technologies.}

Given the relatively limited sample size and following standard practice \citep{xu2022multidimensional}, we first perform marginal logistic regression to select the top 1,000 genes with the smallest $p$-values for downstream analysis. We further incorporate 82 genes from these six prior studies (Table C1), resulting in a total of 1,054 genes in the final dataset. Since pre-classifying prior sources as reliable or not is generally infeasible, BPL, WPL, BP, and WP methods are not applicable. We therefore compare the proposed method with SPL, EWPL and Lasso. $\eta$ and $\lambda$ are selected via 3-fold cross-validation, with $\eta$ tuned over the grid $(0, 0.5, 1, 5, 10, 20, 40, 60)$.

The optimal values of $\eta$ are selected to be 40 in both MPL and EWPL, and 10 in SPL, indicating that integrating multiple priors offers stronger support. SPL assigns full weight to \textit{Prior6}, while MPL distributes weights across \textit{Prior1}-\textit{Prior6}  as 0, 0.248, 0.011, 0.406, 0.090, and 0.244, respectively. This suggests that when several priors are potentially reliable, SPL is limited to selecting only one, whereas MPL effectively identifies and combines all reliable sources. The top-weighted priors identified by MPL -- \textit{Prior2}, \textit{Prior4}, and \textit{Prior6} -- are well-founded. All three share similar research objectives with our study, namely identifying genes associated with TNBC, and partially rely on TCGA data. 

To evaluate the similarity of results across methods, we compute the number of overlapping selected genes between each pair of methods. In addition, we calculate the RV coefficient, which quantifies the similarity between the selected gene expression matrices. As shown in Table \ref{table_rv}, methods that integrate multiple priors (MPL and EWPL) exhibit the highest overlaps and RV coefficients.

\begin{table}
	\caption{Breast cancer data analysis using the proposed and alternative methods: number of overlapping gene identifications (RV-coefficient). } 
	\label{table_rv}
	\begin{center}\small
		\setlength{\tabcolsep}{2mm}{ 
			\begin{tabular}{ccccccc} 
				\hline				
				& MPL & SPL & EWPL  & Lasso & Fwelnet   \\
				 \hline	
				MPL &    29     & 14 (0.938) & 25 (0.977)& 16 (0.953) &13 (0.942)    \\
				SPL &          &  20     & 16 (0.946)      & 14 (0.964) & 12 (0.948)  \\
				EWPL   &          &            &  35      & 18 (0.956)   & 16 (0.948)     \\
				Lasso  &          &            &            & 20          & 15 (0.966)  \\
                Fwelnet &          &            &            &           & 19\\
				\hline	
		\end{tabular}}
\end{center}
\end{table}

\begin{table}\scriptsize
	\caption{Breast cancer analysis: genes selected by MPL, SPL, EWPL, Lasso and Fwelnet methods. A check mark ($\checkmark$) indicates selection by the corresponding method.} 
	\label{table_gene}
		\setlength{\tabcolsep}{0.8mm}{ 
			\begin{tabular}{clccccclccccc} 
				\hline	
&Gene & MPL & SPL & EWPL & Lasso & Fwelnet &   Gene & MPL & SPL & EWPL & Lasso & Fwelnet\\  
\hline  
matched 
& \emph{MLPH} & $\checkmark$  & $\checkmark$ & $\checkmark$ &$\checkmark$ &$\checkmark$
& \emph{CXCL5} & $\checkmark$  & $\checkmark$ & $\checkmark$&$\checkmark$ &$\checkmark$\\
&\emph{ART3} & $\checkmark$  & $\checkmark$ &$\checkmark$& $\checkmark$ &$\checkmark$
&\emph{FOXL1} & $\checkmark$  & $\checkmark$& $\checkmark$& $\checkmark$ &$\checkmark$\\
&\emph{AGR3} & $\checkmark$  & $\checkmark$ &$\checkmark$& $\checkmark$ &$\checkmark$
&\emph{LYPD1} & $\checkmark$  & $\checkmark$ & $\checkmark$& $\checkmark$ &$\checkmark$\\		
&\emph{ESR1} & $\checkmark$  & $\checkmark$ &$\checkmark$& $\checkmark$ &$\checkmark$
&\emph{ERBB2} & $\checkmark$  & $\checkmark$ & $\checkmark$& $\checkmark$ &$\checkmark$\\		
&\emph{POU5F1} & $\checkmark$  & $\checkmark$ &$\checkmark$& $\checkmark$ &$\checkmark$
&\emph{TFF1} & $\checkmark$  & $\checkmark$ &$\checkmark$& $\checkmark$ &$\checkmark$\\	
\hline 
unmatched 
&\emph{C20orf148} & $\checkmark$  & $\checkmark$ &$\checkmark$& $\checkmark$ & 	
&\emph{DEGS2 } &   &$\checkmark$ & $\checkmark$&\\
& \emph{AK8} & $\checkmark$  & & $\checkmark$ &$\checkmark$	&$\checkmark$	
& \emph{ANKRD30A} &   &$\checkmark$ &  &$\checkmark$ & \\
& \emph{LYAR} & $\checkmark$  &  & $\checkmark$&  $\checkmark$&$\checkmark$
& \emph{FSIP1} &   & & $\checkmark$ & &$\checkmark$\\
& \emph{LYPD5} & $\checkmark$  & &$\checkmark$ &  $\checkmark$ &$\checkmark$
& \emph{CHODL} &  $\checkmark$ & &  & &\\
& \emph{FZD9} &   &$\checkmark$ & $\checkmark$& $\checkmark$ &$\checkmark$
& \emph{DKC1 } &  $\checkmark$ & &  &\\
& \emph{LINC02872} &   &$\checkmark$ & $\checkmark$&  $\checkmark$&$\checkmark$
&\emph{IFRD1} &$\checkmark$ & & & &\\
& \emph{ARL9} & $\checkmark$  & $\checkmark$ & $\checkmark$& &		
&\emph{FOXA1 } &   &$\checkmark$ & &\\
& \emph{LMX1B} & $\checkmark$  & $\checkmark$ & $\checkmark$& &
&\emph{C9orf152} &   &$\checkmark$ & &\\ 
& \emph{KCNS1} & $\checkmark$  & & $\checkmark$&  $\checkmark$ &
&\emph{UCHL1} & & &$\checkmark$ & &\\
& \emph{FDCSP} & $\checkmark$  &  & $\checkmark$&  $\checkmark$  &
&\emph{CCKBR} & & &$\checkmark$ & &\\
& \emph{CSNK1A1} & $\checkmark$  & $\checkmark$ & & &
&\emph{PIF1} & & &$\checkmark$ & &\\
& \emph{IGF2BP3 } &  $\checkmark$ & &$\checkmark$ & &
&\emph{PCDH8} &   & & $\checkmark$ & & \\
&\emph{ENPP1 } &  $\checkmark$ & &$\checkmark$ & &
& \emph{GFRA3} &   & & $\checkmark$ & &\\
&\emph{C9orf40 } &  $\checkmark$ & & $\checkmark$ & &
&\emph{ADD2} & & &$\checkmark$ & &\\
&\emph{IL22RA2 } &  $\checkmark$ & &$\checkmark$& &
&\emph{CT62} &   & & &$\checkmark$  &\\
& \emph{FAM19A4 } &  $\checkmark$ & &$\checkmark$ & &
&\emph{CT83} & & & & & $\checkmark$\\
&\emph{CNKSR2 } &  $\checkmark$ & &$\checkmark$ & &
& \emph{BCAS1} &   & & & &$\checkmark$\\
& \emph{AFAP1-AS1 } &  $\checkmark$ & &$\checkmark$ & &
&\emph{CA12} & & & & & $\checkmark$\\
				\hline
		\end{tabular}}
\end{table}

Table \ref{table_gene} summarizes the gene selection results across all four methods. A total of 10 genes are identified by all methods, while 36 genes are absent in at least one. MPL selects 29 genes, including three -- \textit{CHODL},  \textit{DKC1} and \textit{IFRD1} -- that are uniquely identified. Notably, these genes have been previously implicated in breast cancer. For instance, 
\textit{CHODL} is potentially implicated in tumor metastasis, participating in cellular recognition, communication, cell-to-cell adhesion, and extracellular matrix interactions \citep{masuda2011chondrolectin}.
\textit{DKC1} is overexpressed in TNBC and linked to poor clinicopathological features and unfavorable prognosis \citep{vilarullo2025improving}. 
The m6A methylation of \textit{IFRD1} RNA affects the proliferation of breast cancer cells \citep{zhang2024mthfd2}.
These findings support the ability of MPL to uncover biologically relevant signals missed by alternative approaches.

To evaluate predictive performance, we conduct a random splitting-based assessment. Specifically, the data is randomly split into training and testing sets with a 2:1 ratio. Models are trained on the training set, and prediction misclassification rates (PMRs) are evaluated on the testing set. MPL achieves the best performance with a PMR (sd) of 0.056 (0.009), followed by SPL at 0.061 (0.007), EWPL at 0.060 (0.015), Lasso at 0.094 (0.022), and the Fwelnet at 0.155 (0.040). 

\section{Discussion}\label{Discussion}

This study has developed the adaptive Multi-Prior Lasso, an important approach for incorporating multiple sources of priors into high-dimensional GLMs when individual-level data from external studies are unavailable. By adaptively evaluating the concordance between each prior and the current data, the method assigns informative weights and selectively integrates reliable prior information to improve model estimation. Theoretical analysis establishes consistency and selection properties, while simulation results demonstrate superior estimation, variable selection, and predictive accuracy compared to the existing approaches. Through genetic analysis of TNBC, the method has successfully identified biologically meaningful signals -- including genes uniquely recovered by MPL -- that are supported by prior literature but missed by the alternatives. Results highlight the practical advantage of leveraging multiple priors.

This work opens several avenues for future research. Extending the approach to handle survival outcomes, such as in Cox models, can broaden its utility. Robustness against outliers and contamination, as well as adaptations to account for missing data may be next steps. More broadly, the proposed approach offers a novel and scalable paradigm for integrating prior knowledge in high-dimensional studies, beyond the GLM framework, and is particularly appealing when prior findings are abundant but access to individual-level data is constrained.

\section*{Acknowledgements}
This study is supported by the National Natural Science Foundation of China (12571292), NIH (CA204120), NSF (2209685), and the China Postdoctoral Science Foundation (2025M773107). Fuzhi Xu and Weijuan Liang contributed equally.

\bibliographystyle{ECA_jasa}

\end{document}